\documentclass[12pt]{article}

\usepackage{amsmath,amsfonts}
\usepackage{amssymb}
\usepackage{hhline}
\usepackage{epsfig,cite}
\usepackage{stmaryrd}
\usepackage[usenames,dvips]{xcolor}

\usepackage{fullpage}
\usepackage{verbatim}
   \allowdisplaybreaks
\makeatletter
\@addtoreset{equation}{section}
\makeatother

\newcommand{\be}{\begin{equation}}
\newcommand{\ee}{\end{equation}}
\newcommand{\bea}{\begin{eqnarray}}
\newcommand{\eea}{\end{eqnarray}}

\definecolor{red}{rgb}{1,0,0}

\begin{document}

\thispagestyle{empty}

\begin{center}
\hfill CERN-PH-TH/2013-301\\
\hfill UAB-FT-749\\
\hfill NSF-KITP-13-258

\begin{center}

\vspace{.5cm}

{\Large\sc Electroweak and supersymmetry breaking\\ \vspace{0.4cm}from the Higgs discovery}

\end{center}

\vspace{1.cm}

\textbf{ Antonio Delgado$^{\,a,\,b}$, Mateo Garcia$^{\,c}$,
and Mariano Quiros$^{b,\,d,\,e}$}\\

\vspace{1.cm}
${}^a\!\!$ {\em {Department of Physics, University of Notre Dame, Notre Dame, IN 46556, USA}}

\vspace{.1cm}
${}^b\!\!$ {\em {KITP, University of California, Santa Barbara, CA 93106-4030, USA}}

\vspace{.1cm}
${}^c\!\!$ {\em {IFAE, Universitat Aut{\`o}noma de Barcelona,
08193 Bellaterra, Barcelona, Spain}}

\vspace{.1cm}

${}^d\!\!$ {\em {Department of Physics, CERN-TH Division, CH-1211, Geneva 23, Switzerland}}

\vspace{.1cm}
${}^e\!\!$ {\em
{Instituci\'o Catalana de Recerca i Estudis  
Avan\c{c}ats (ICREA) and\\ Institut de F\' isica d'Altes Energ\`ies, UAB,
08193 Bellaterra, Barcelona, Spain}}

\end{center}

\vspace{0.8cm}

\centerline{\bf Abstract}
\vspace{2 mm}
\begin{quote}\small
We will explore the consequences on the electroweak breaking condition, the mass of supersymmetric partners and the scale at which supersymmetry breaking is transmitted, for arbitrary values of the supersymmetric parameters $\tan\beta$ and the stop mixing $X_t$, which follow from the Higgs discovery with a mass $m_H\simeq 126$ GeV at the LHC. Within the present uncertainty on the top quark mass we deduce that radiative breaking requires $\tan\beta \gtrsim 8$ for maximal mixing $X_t\simeq \sqrt{6}$, and $\tan\beta\gtrsim 20$ for small mixing $X_t\lesssim 1.8$. The scale at which supersymmetry breaking is transmitted $\mathcal M$ can be of order the unification or Planck scale only for large values of $\tan\beta$ and negligible mixing $X_t\simeq 0$. On the other hand for maximal mixing and large values of $\tan\beta$ supersymmetry should break at scales as low as $\mathcal M\simeq 10^5$ GeV. The uncertainty in those predictions stemming from the uncertainty in the top quark mass, i.e.~the top Yukawa coupling, is small (large) for large (small) values of $\tan\beta$. In fact for $\tan\beta=1$ the uncertainty on the value of $\mathcal M$ is several orders of magnitude.

 \end{quote}

\vfill

 \newpage

\section{Introduction and summary}
After the 7 and 8 TeV runs the Large Hadron Collider (LHC) has firmly established the existence of a scalar boson with a mass $m_H\simeq 126$ GeV. In particular the strengths measured in the $WW$, $ZZ$, $\gamma\gamma$, $b\bar b$ and $\tau\tau$ decay channels by the ATLAS and CMS collaborations are consistent with the Standard Model (SM) Higgs with a mass $m_H=126\pm 0.4 \textrm{ (stat)} \pm 0.4 \textrm{ (syst)}$ GeV~\cite{Aad:2012tfa} and $m_H=125.3\pm 0.4 \textrm{ (stat)} \pm 0.5 \textrm{ (syst)}$ GeV~\cite{Chatrchyan:2012ufa}, respectively. The Higgs discovery is of the upmost importance as it is the first direct experimental confirmation of the mechanism of electroweak (EW) symmetry breaking (EWSB). In the SM it points toward a quartic coupling $\lambda=m_H^2/v^2$, where $v=246$ GeV, provided by the Higgs potential 
\be
V_{\textrm SM}=-m^2|H|^2+\frac{\lambda}{2}|H|^4
\ee
where $m^2=\lambda v^2/2$, valid at the EW scale $\mathcal Q_{EW}=m_H$.

From the theoretical point of view we know that the the EW minimum is unstable against quantum corrections (a problem known in the literature as the hierarchy problem) and has to be stabilized by some beyond the SM (BSM) physics, the paradigm of which being supersymmetry and in particular the minimal SM supersymmetric extension (MSSM). Nonetheless another feature of the past LHC runs is that no experimental hints have been found of BSM particles which could stabilize the EW vacuum, but it is putting bounds on the mass of supersymmetric particles~\cite{susyexp}. Still, and in view of the forthcoming LHC run at 13-14 TeV, it is interesting to explore the consequences of the present Higgs mass data on a possible underlying supersymmetric theory, in particular on the way supersymmetry triggers EWSB at low energy and on the value of the scale at which supersymmetry is broken.

In this paper we will then consider at face value the present data on the Higgs sector. We will assume that the SM emerges at some scale $\mathcal Q_0$ from an underlying MSSM, and will extract the relevant information on the mechanism by which the MSSM triggers EWSB and on the scale $\mathcal M$ at which  supersymmetry is broken in the hidden sector. Consistently with present experimental data we will assume that below the scale $\mathcal Q_0$ we just have the SM spectrum and the matching conditions are the ones to enforce EWSB at the EW scale $\mathcal Q_{EW}=m_H$.

 The contents of this paper are as follows. In Sec.~\ref{matching} we impose the condition that the SM and the MSSM merge at the scale $\mathcal Q_0$ and compute it by fixing the Higgs mass in the SM and with fixed values of $\tan\beta$ and the stop mixing $X_t$ in the MSSM. We see, not unexpectedly, that for low values of $\tan\beta$, $\mathcal Q_0$ is large and insensitive to the mixing $X_t$ while for large values of $\tan\beta$ it can be small and sensitive to $X_t$. In particular, values of $\mathcal Q_0$ in the TeV range require both large values of $\tan\beta$ $(\tan\beta\gtrsim 5)$ and of the mixing $(X_t\gtrsim 1.8)$. Moreover we can translate the condition of EWSB in the SM to a condition on $m_2^2(\mathcal Q_0)$ (the squared mass of the Higgs doublet that gives a mass to the top quark) and so we can scrutinize on the nature of EWSB, i.e.~radiative versus non radiative breaking~\footnote{We will conventionally dub radiative breaking the situation where $m^2(\mathcal Q_0)\leq 0$ although electroweak breaking is triggered in all cases by radiative corrections.}. We have found that the nature of EWSB strongly depends on both parameters, $\tan\beta$ and $X_t$. In particular we have found that radiative breaking requires $\tan\beta\gtrsim 8$ for maximal mixing and $X_t\gtrsim 1.8$ for $\tan\beta\lesssim 20$. Another interesting feature we have found, both in the calculation of the matching scale, i.e.~$\mathcal Q_0$, and the type of EWSB, i.e.~the value of $m_2^2(\mathcal Q_0)$, is that the main uncertainty in the calculation comes from the uncertainty in the top quark mass and that it affects mainly low values of $\tan\beta$ (it is mostly insensitive to the actual value of the mixing $X_t$). In particular for $\tan\beta=1$ and $X_t=0$, while for the central value of the top quark mass $\mathcal Q_0\sim 10^{11}$ GeV, after inserting the 2$\sigma$ uncertainty it can vary in the range $\mathcal Q_0\in[10^{9},10^{16}]$ GeV and no sharp prediction can be made. Notice that the results in this section imply precise values of $m_1(\mathcal Q_0)$ and $m_2(\mathcal Q_0)$ determined by the EWSB condition and by the condition of decoupling of the heavy Higgs at the matching scale $\mathcal Q_0$.

In Sec.~\ref{susybreaking} we  computed the scale at which supersymmetry is transmitted $\mathcal M$ by imposing the condition that the MSSM Higgs mass parameters are equal at that scale: $m_1(\mathcal M)=m_2(\mathcal M)$. All of our results in this section are based on this assumption. This one is a natural assumption in most existing models of supersymmetry breaking, including those coming from string theories. Of course, should the condition on $m_1(\mathcal M)$ and $m_2(\mathcal M)$ be changed our results would correspondingly be modified. To compute the value of the scale at which supersymmetry is broken $\mathcal M$ we have first followed a bottom-up approach where we assume the ideal conditions that all supersymmetric particles decouple exactly at the scale $\mathcal Q_0$ (with no thresholds). In the second, top-down, approach we have instead assumed that the supersymmetric parameters are the ones obtained in various models of supersymmetry breaking transmission to the observable sector. In these cases there are different thresholds around the matching scale $\mathcal Q_0$ but the results are in all cases consistent with the first approach. The main result in the bottom-up approach is (not unexpectedly) that the value of $\mathcal M$ depends to a large extent on the supersymmetric parameters $(\tan\beta(\mathcal Q_0),X_t(\mathcal Q_0))$. In particular for $X_t\geq 0$ large values of $\mathcal M$ close to the unification or Planck scale can only be obtained for large values of $\tan\beta$ and small mixing. Of course for those small values of $X_t$ the constraint on the Higgs mass impose large values of $\mathcal Q_0$, say in the 10 -100 TeV region for which the supersymmetric spectrum would be outside the reach of LHC. On the contrary for large values of $X_t$, for instance for maximal mixing, for which $\mathcal Q_0$ is in the TeV range and the supersymmetric spectrum is inside the reach of LHC, the scale at which supersymmetry breaking is transmitted can go down to the low scales such that gravity mediation mechanisms are precluded. For negative values of $X_t$ there is room for large GUT or Planckian values of $\mathcal M$ provided that $X_t$ is in some intermediate region, e.g.~$X_t\simeq -1.5$ which can accommodate lower values of $\mathcal Q_0$ for large values of $\tan\beta$, e.g. $\mathcal Q_0=\mathcal O$(few) TeV, inside the LHC reach. Again our predictions are affected by the top quark mass uncertainty $\Delta\bar m_t(m_t)$. As it was the case for the $\mathcal Q_0$ prediction, the uncertainty affects mainly small values of $\tan\beta$ and it is rather insensitive to the value of $X_t$. In the second, top-down, approach we have considered two different cases where supersymmetric parameters unify at the scale $\mathcal M$. First we have considered  the case of universal soft parameters, by which all squark masses ($m_0$), all gaugino masses ($m_{1/2}$) and all Higgs mass parameters ($m_H$) unify at the scale $\mathcal M$. This is a general CMSSM where we have separated the Higgs from the sfermion masses and which can appear in gravity mediated supersymmetry breaking theories. We have considered two examples with $\tan\beta=10$ and $X_t=0,\,2$. In agreement with the results of the previous section the case $X_t=0$ is consistent with EWSB and the Higgs mass $m_H=126$ GeV for $\mathcal M\simeq 10^{18}$ GeV, while the case $X_t=2$ requires  supersymmetry breaking at low scale $\mathcal M\simeq 10^{6}$ GeV, hard to reconcile with gravity mediation. The second case we have considered is the minimal gauge mediated supersymmetry breaking (GMSB) where the mass of scalars transforming under a gauge group $G_a$, with gauge coupling $\alpha_a$, and the corresponding gaugino is proportional to $\alpha_a(\mathcal M)/4\pi$ and the trilinear coupling is $A_t(\mathcal M)=0$. Below $\mathcal M$, $A_t$ is generated by the MSSM RGE and therefore it gets negative values at the scale $\mathcal Q_0$, giving then $X_t<0$. We have presented two cases with $N=4$ messengers, $\tan\beta=[15,\, 8]$ and values of $\mathcal M=[10^{8},\, 10^{11}]$ GeV and $X_t=[-1.8,-1.6]$ which are consistent with perturbative unification at the MSSM GUT scale.
Finally in Sec.~\ref{conclusion} we present our conclusions.

\section{The matching and electroweak breaking}
\label{matching}
The quadratic terms in the MSSM potential can be written as
\be
V_2=m_1^2|H_1|^2+m_2^2|H_2|^2+m_3^2(H_1\cdot H_2+h.c.)
\ee
 with $H_1\cdot H_2\equiv H_1^a\varepsilon_{ab}H_2^b$ $(\varepsilon_{12}=-1)$ and we are defining $m_1^2=m_{H_1}^2+\mu^2$ and $m_2^2=m_{H_2}^2+\mu^2$, where $m_{H_i}$ is the  soft breaking mass for $H_i$ and $\mu$ is the supersymmetric Higgsino mass. They can also be written as
 \be
 V_2=(H_1^\dagger,\widetilde H_2^\dagger) \left(\begin{array}{cc}m_1^2 &m_3^2\\m_3^2 & m_2^2\end{array}\right)
 \left(\begin{array}{c}H_1\\\widetilde H_2\end{array}\right)
 \ee
 where $\widetilde H_2\equiv \varepsilon H_2^*$. The diagonalization of the mass matrix 
 \begin{equation}
 \mathcal M_0^2=\begin{pmatrix} m_1^2 & m_3^2\\ m_3^2 & m_2^2\end{pmatrix}
 \end{equation}
 then yields the mass eigenvalues
 \begin{equation}
 m^2_{\mp}=\frac{m_1^2+m_2^2}{2}\mp \sqrt{\left(\frac{m_1^2-m_2^2}{2}\right)^2+m_3^4}
 \end{equation}
\subsection{The matching scale} 
 We wish to match the MSSM with the SM at the (common) scale $\mathcal Q_0\equiv m_0$ of supersymmetric masses. In particular we will rotate the MSSM Higgs sector $(H_1,\widetilde H_2)$ into the basis $(H,\mathcal H)$ where $H$ is the SM Higgs doublet and $\mathcal H$ its heavy orthogonal combination. We then identify the mass squared of the (light) SM Higgs $H$ with the tachyonic mass $m_-^2= -m^2(\mathcal Q_0)$  and consequently the mass squared of its (heavy) orthogonal combination $\mathcal H$ with $m_{+}^2\equiv m_{\mathcal H}^2=m_1^2+m_2^2+m^2$. This can be done by the fixing
 \begin{equation}
 m_3^4=(m_1^2+m^2)(m_2^2+m^2)
 \end{equation}
 leading to the mixing angle $\beta$ given by
 \begin{equation}
 \tan^2\beta=\frac{m_1^2+m^2}{m_2^2+m^2}\quad \text{i.e.}\quad
 m^2=\frac{m_1^2-m_2^2\tan^2\beta}{\tan^2\beta-1}
 \label{minimum}
 \end{equation}
 where all quantities are evaluated at the matching scale $\mathcal Q=\mathcal Q_0$, which rotates the Higgs basis $(H_1,\widetilde H_2)$ into the mass eigenstates $(H,\mathcal H)$ as
 \bea
 H&=&\cos\beta H_1-\sin\beta \widetilde H_2\nonumber\\
 \mathcal H&=&\sin\beta H_1+\cos\beta \widetilde H_2\ .
 \eea
 The potential for the SM Higgs then reads as
 \be
 V_{\rm SM}=-m^2(\mathcal Q_0)|H|^2+\frac{\lambda(\mathcal Q_0)}{2}|H|^4+\cdots
 \label{SMpot}
 \ee

 In order to make a precise calculation of the Higgs mass we have to first match the SM quartic coupling $\lambda$ and the supersymmetric parameters at the scale $\mathcal Q_0$.  We will improve over the tree-level ($\ell=0$) matching by considering the one-loop $(\ell=1)$ and leading two-loop ($\ell=2$) threshold effects as given by~\cite{Draper:2013oza}
 \be
   \lambda(\mathcal Q_0)=\sum_{\ell\ge 0}\Delta^{(\ell)}\lambda
  \ee 
 where 
 \begin{align}
 \Delta^{(0)}\lambda&=\frac{1}{4}(g^2+g^{\prime\,2})c^2_{ 2\beta}\nonumber\\
 16\pi^2 \Delta^{(1)}\lambda&=6 y_t^4s_\beta^4 X_t^2\left(1-\frac{X_t^2}{12}\right)-\frac{1}{2}y_b^4s_\beta^4(\mu/\mathcal Q_0)^2+\frac{3}{4}y_t^2 s_\beta^2(g^2+g'^2)X_t^2c_{2\beta}\nonumber\\
 &+\left(\frac{1}{6}c^2_{2\beta}-\frac{3}{4} \right)g^4-\frac{1}{2}g^2g'^2-\frac{1}{4}g'^4-
 \frac{1}{16}(g^2+g'^2)^2s^2_{4\beta}\nonumber\\
  (16\pi^2)^2 \Delta^{(2)}\lambda&=16 y_t^4 s_\beta^4 g_3^2\left(-2X_t+\frac{1}{3}X_t^3-\frac{1}{12}X_t^4\right)+\mathcal O(h_t^6s_\beta^4,\, g^4,\,g^2g'^2,\,g'^4)
 \label{lambda}
 \end{align}
and we are using the notation $X_t=(A_t(\mathcal Q_0)-\mu(\mathcal Q_0)/\tan\beta)/\mathcal Q_0$, and $s_\beta\equiv \sin \beta$ and so on. For the numerical calculation we are also taking into account the $\mathcal O(y_t^6s_\beta^4,\dots)$ two-loop threshold corrections whose explicit expression can be found in Ref.~\cite{Draper:2013oza}. We are neglecting the corrections proportional to $y_\tau^4$ as we are not envisaging values of the parameter $\tan\beta$ such that $y_\tau$ is relevant.

The couplings $y_t$ and $y_b$ are the top and bottom Yukawa couplings in the MSSM. They are related to the corresponding SM couplings $h_t$ and $h_b$ by~\cite{Draper:2013oza}
\begin{align}
h_{t}&=y_t s_\beta\left(1-\frac{1}{6\pi^2}g_3^2 \mathcal Q_0^2 X_t I(m_{\tilde t_1},m_{\tilde t_2},\mathcal Q_0) +\mathcal O(y_b^2,\,g^2,\,g'^2)) \right)\nonumber\\
h_{b}&=y_b c_\beta\left(1-\frac{1}{6\pi^2}g_3^2 \mathcal Q_0^2 X_b I(m_{\tilde b_1},m_{\tilde b_2},\mathcal Q_0) +\frac{1}{16\pi^2}y_t^2 t_\beta \mathcal Q_0^2 X_t I(m_{\tilde t_1},m_{\tilde t_2},\mathcal Q_0)+\dots \right)
\end{align}
where $X_b=(A_t(\mathcal Q_0)-\mu(\mathcal Q_0)\tan\beta)/\mathcal Q_0$, we are assuming nearly degenerate spectrum at $\mathcal Q_0$, and only the leading one-loop QCD and top Yukawa coupling corrections are kept. The function $I(x,y,z)$ can be found in Ref.~\cite{Draper:2013oza}. 

The parameters of the potential (\ref{SMpot}) have to be run with the SM RGE down to the  scale $\mathcal Q_{EW}=m_H$, where minimizing the SM potential should lead to $m^2(m_H)=\frac{1}{2}m_H^2$, $m_H^2=2\lambda(m_H) v^2$. For a similar analysis see Ref.~\cite{Giudice:2006sn}. Here in agreement with the used threshold corrections we are using the two-loop RGE as 
given in~\cite{Arason:1991ic}.

Finally going from the running Higgs mass $m_H$ to the pole Higgs mass $M_H$ requires the calculation of the Higgs boson self energy $\Pi(p^2)$ as $M_H^2=m_H^2+\Delta\Pi$ where $\Delta\Pi=\Pi(p^2=M_H^2)-\Pi(p^2=0)$. Here we keep only the leading correction to $\Delta\Pi$ coming from the top quark loop exchange given by~\cite{Casas:1994us}
\be
\Delta\Pi_{tt}=\frac{3h_t^2 M_t^2}{4\pi^2}\left[2-Z(M_t^2/M_H^2)\right],\ Z(x)=2\sqrt{4x-1}\arctan \left(1/\sqrt{4x-1}\right),\ x>1/4
\label{pole}
\ee
 For the actual values of $M_t\simeq\overline{m}(m_t)+10$ GeV~\cite{Alekhin:2012py} (the pole top quark mass) and $M_H$, the correction in (\ref{pole}) is of the order of the experimental error in the Higgs mass.
 
\begin{figure}[htb]
\begin{center}
\includegraphics[width=80.mm,height=70mm]{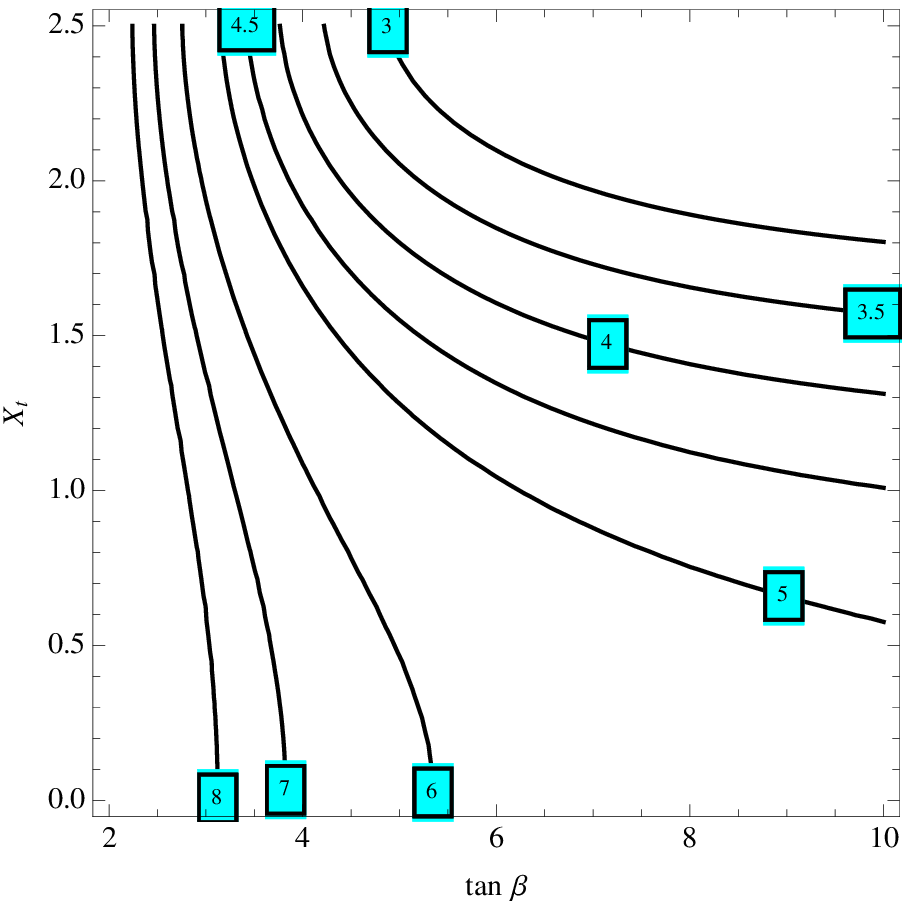}
\includegraphics[width=80.mm,height=70mm]{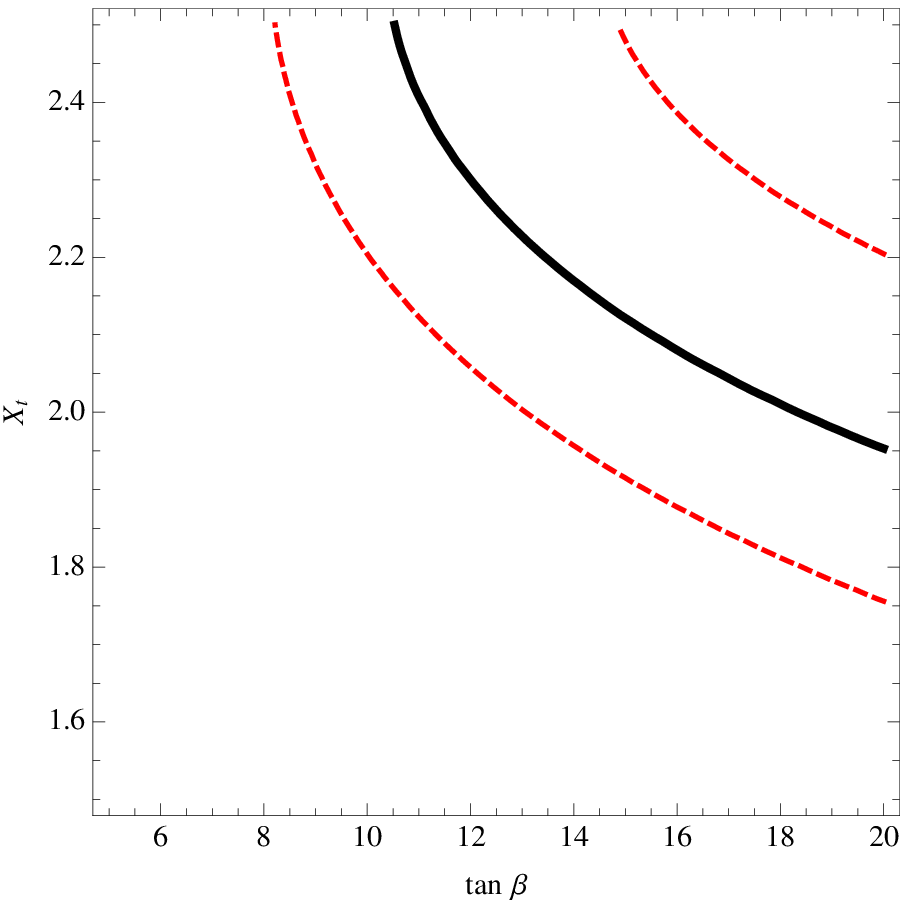}
\end{center}
\caption{\it  Left panel: Contour lines of $\log_{10}[ \mathcal Q_0/\text{GeV}]$ (for the values specified in the plot) in the plane $(\tan\beta,X_t)$. Right panel: Contour line of $m_2^2(\mathcal Q_0)=0$, as given by Eq.~(\ref{breaking}), in the plane $(\tan\beta,X_t)$. The inner region corresponds to radiative electroweak breaking.}
\label{Q0contours}
\end{figure}

Notice that, for fixed values of the supersymmetric parameters $\tan\beta$ and $X_t$, $\mathcal Q_0$ is a function of the Higgs mass $m_H$. This prediction comes from the intersection of the function $\lambda(\mathcal Q)$, which is determined mainly by the value of the Higgs mass [with some dependence on the actual values of $h_t(m_H)$ and $\alpha_3(m_H)$], with the value $\lambda(\mathcal Q_0)$ given by Eq.~(\ref{lambda}). So given that the Higgs mass is fixed to $m_H=126$ GeV, we can predict $\mathcal Q_0=\mathcal Q_0(\tan\beta,X_t)$ as it is shown in the left panel plot of Fig.~\ref{Q0contours}.
 We have used as an input the running top mass in the $\overline{\rm MS}$ scheme evaluated at the top mass $\overline{m}_t(m_t)=163.5$ GeV. We can see that for small values of $\tan\beta$ the values of $\mathcal Q_0$ are large and insensitive to the values of the mixing $X_t$. This is due to the fact that the threshold effect is proportional to $h_t^2(\mathcal Q_0)$ and the Standard Model RGE leads to small values of $h_t(\mathcal Q_0)$ for large values of the scale $\mathcal Q_0$. On the other hand for large values of $\tan\beta$ the values of $\mathcal Q_0$ are smaller and consequently the RGE running is small and $\mathcal Q_0$ becomes sensitive to the mixing $X_t$.  In particular values of $\mathcal Q_0$ in the TeV region require large values of $\tan\beta$ ($\tan\beta\gtrsim 5$) and large values of $X_t$ ($X_t\gtrsim 1.8$).
 
\begin{figure}[htb]
\begin{center}
\includegraphics[width=81.5mm]{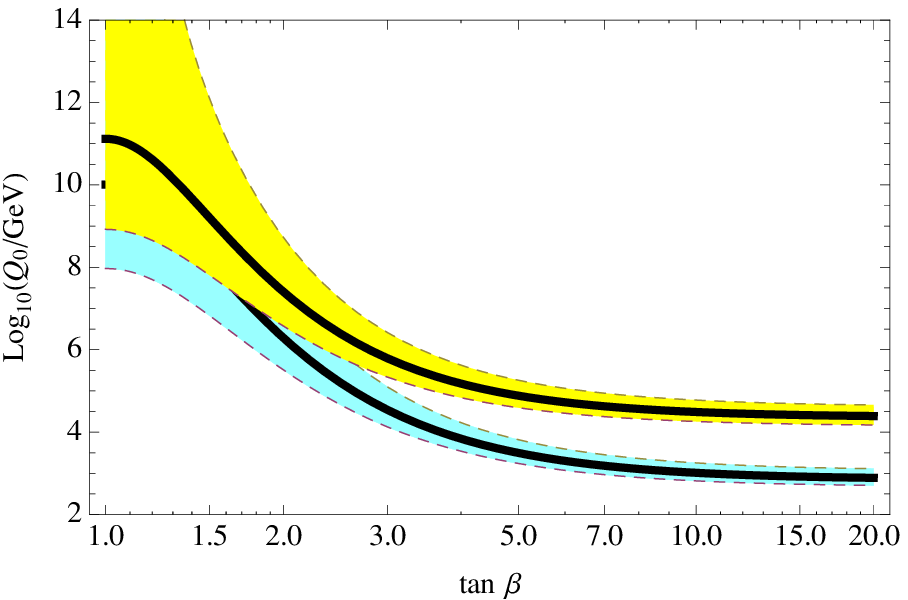}
\includegraphics[width=81.5mm]{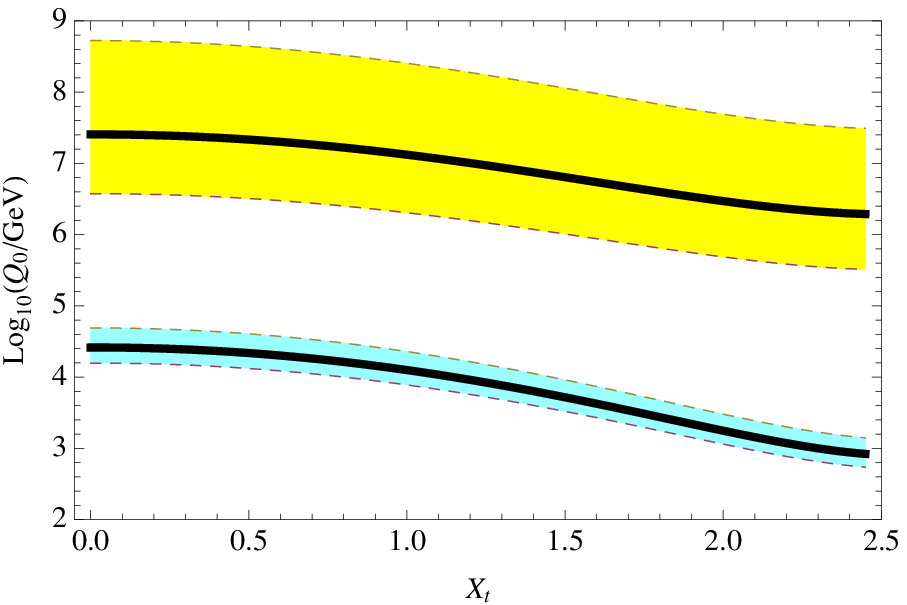}
\end{center}
\caption{\it Left panel: Plot of $\mathcal Q_0$ as a function of $\tan\beta$ for $X_t=$0 (upper band) and $X_t=\sqrt{6}$ (lower band). The width of bands corresponds to the experimental error $\Delta \overline{m}_t=\pm 2$ GeV. Right panel:  Plot of $\mathcal Q_0$ as a function of $X_t$ for $\tan\beta=2$ (upper band) and $15$ (lower band). }
\label{Q0}
\end{figure}

 As for the error in $\overline{m}_t(m_t)$ it is safe to consider the experimental range of the running top mass to be given by $\Delta \overline{m}_t=\pm 2$ GeV at 2$\sigma$~\cite{Alekhin:2012py,Masina:2012tz}. In order to see the relevance of the error in $\overline{m}_t(m_t)$ we plot, in the left panel of Fig.~\ref{Q0}, $\mathcal Q_0$ as a function of $\tan\beta$ for various values of $X_t$, and in the right panel of Fig.~\ref{Q0},
 $\mathcal Q_0$ as a function of $X_t$ for different values of $\tan\beta$.
In fact the upper border of each band corresponds to $\Delta\overline{m}_t=-2$ GeV and the lower border to $\Delta \overline{mÐ}_t=+2$ GeV.
We can see from both panels of Fig.~\ref{Q0} that the error in the determination of $\mathcal Q_0$, $\Delta\mathcal Q_0$ arising from the error in $\overline{m}_t(m_t)$ is large (small) for small (large) values of $\tan\beta$. The reason for this behavior is that the error $\overline{m}_t(m_t)$ is amplified by the RGE running and it is consequently large (small) for large (small) running, which means small (large) values of $\tan\beta$. In the same way, as we can see from the right panel of Fig.~\ref{Q0}, the error $\Delta\mathcal Q_0$ is uncorrelated with $X_t$ as it has little influence on the RGE running. This translates into a big overlapping in the left panel of Fig.~\ref{Q0} for small values of $\tan\beta$ and different values of $X_t$. In fact notice that for the limiting case $\tan\beta=1$ and $X_t=0$ we have that $\lambda(\mathcal Q_0)\lesssim 0$ and the Standard Model potential is unstable. This corresponds, for the central value of the quark top mass, to $\mathcal Q_0\sim 10^{11}$ GeV. However for the lowest allowed value of the top quark mass the instability scale can go to Planckian values in agreement with various calculations in the literature~\cite{Degrassi:2012ry,Buttazzo:2013uya}. In this case it has been shown that the Veltman condition~\cite{Veltman:1980mj} (or absence of quadratic divergences) can also be satisfied~\cite{Masina:2013wja}.  

\subsection{Electroweak breaking}
As we have noticed  Eq.~(\ref{minimum}) actually implies the existence of the electroweak minimum in the SM effective theory and indeed it is reminiscent of the minimum equation in the MSSM~\footnote{Were we neglecting the Standard Model RGE running both equations would be equivalent upon identification of $m_H^2\leftrightarrow m_Z^2$.}. In fact Eq.~(\ref{minimum}) can be traded by the SM minimum equation. It can be written as 
 \be
 m_2^2(\mathcal Q_0)=\frac{m_{\mathcal H}^2(\mathcal Q_0)-m^2(\mathcal Q_0)\tan^2\beta}{\tan^2\beta+1}
\label{breaking}
 \ee
where we identify $m_{\mathcal H}^2(\mathcal Q_0)\equiv \mathcal Q_0^2$ and the value obtained for $m_2^2(\mathcal Q_0)$  characterizes the type of electroweak breaking, e.g.~radiative versus non-radiative symmetry breaking~\footnote{Although EW breaking is in all cases driven by the MSSM RGE running from $\mathcal M$ to $\mathcal Q_0$, we will be conventionally dubbing radiative breaking the case where $m_2^2(\mathcal Q_0)\leq 0$ so that the EW breaking proceeds by a tachyonic mass as in the SM.}, provided that after the SM RGE running we get $m^2(\mathcal Q_{EW})=m_H^2/2$. For instance in the limit $\tan\beta\to\infty$ [or more precisely for $\tan^2\beta\gg m_\mathcal H^2(\mathcal Q_0)/m^2(\mathcal Q_0)$] we get the conditions for radiative breaking, $m_2^2(\mathcal Q_0)\simeq -m^2(\mathcal Q_0)<0$, while for small values of $\tan\beta$ we get the conditions for non-radiative breaking $m_2^2(\mathcal Q_0)\simeq {\displaystyle \frac{m_{\mathcal H}^2(\mathcal Q_0)}{\tan^2\beta+1} }>0$. In particular we show in the right panel of Fig.~\ref{Q0contours} the contour plot corresponding to $m_2^2(\mathcal Q_0)=0$ for the central value of $\overline{m}_t(m_t)$ (thick solid line) and for the $2\sigma$ values corresponding to $\pm \Delta\overline{m}_t(m_t)$ (thin solid lines). The inner area corresponds to the region where there is radiative electroweak symmetry breaking $m_2^2(\mathcal Q_0)<0$ while in the outer region the breaking is not radiative and  $m_2^2(\mathcal Q_0)>0$. Of course the values of $m_2^2(\mathcal Q_0)$ should depend to a large extent on the values of $\tan\beta$ and $X_t$.
 \begin{figure}[htb]
\begin{center}
\includegraphics[width=81.5mm]{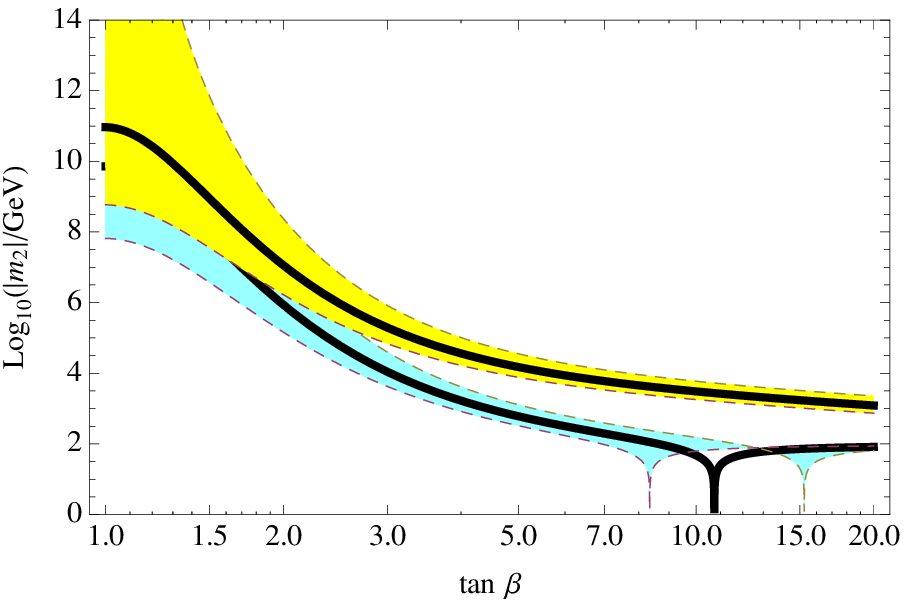}
\includegraphics[width=81.5mm]{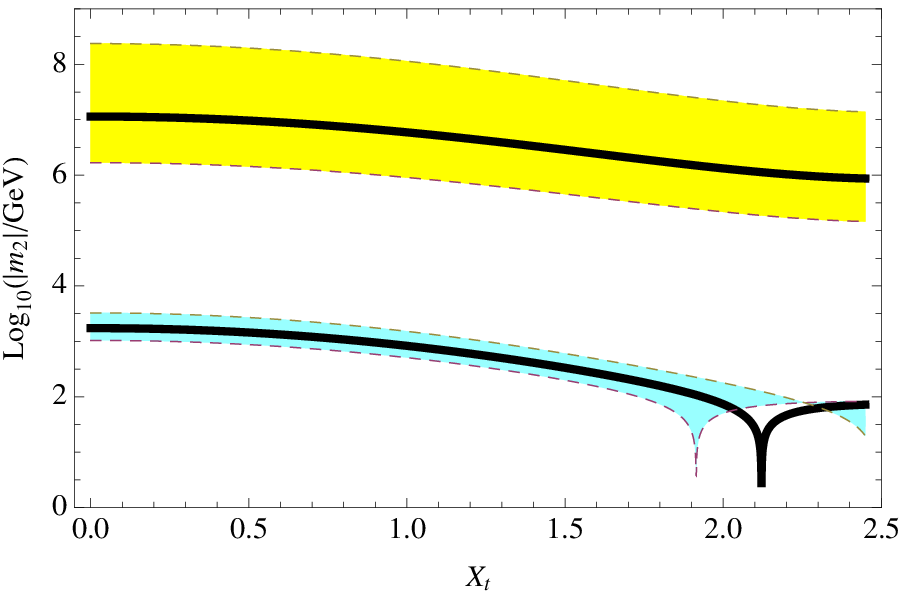}
\end{center}
\caption{\it Left panel: Plot of $|m_2(\mathcal Q_0)|$ as a function of $\tan\beta$ for $X_t=$0 (upper band) and $X_t=\sqrt{6}$ (lower band). The width of bands corresponds to the experimental error $\Delta \overline{m}_t=\pm 2$ GeV. Right panel:  Plot of $|m_2(\mathcal Q_0)|$ as a function of $X_t$ for $\tan\beta=2$ (upper band) and $15$ (lower band). }
\label{m2}
\end{figure}

In Fig.~\ref{m2} we plot the absolute value of $m_2$, $|m_2(\mathcal Q_0)|$, as a function of $\tan\beta$ for different values of $X_t$ (left panel) and as a function of $X_t$ for different values of $\tan\beta$ (right panel). Notice that points where electroweak breaking becomes radiative are characterized by the fact that $|m_2|=0$ and for larger values of $\tan\beta$ (left panel of Fig.~\ref{m2}) or larger values of $X_t$ (right panel of Fig.~\ref{m2}), $m_2^2$ becomes negative and thus $|m_2|$ takes on positive values. Again we can see that, as for the results in Fig.~\ref{Q0}, the effects of the error $\Delta\overline{m}_t(m_t)$ are amplified for small values of $\tan\beta$ while they stay small for large values of $\tan\beta$. We can also see that radiative breaking only occurs for large values of $\tan\beta$, $\tan\beta\gtrsim 8$, and/or large values of the mixing $X_t\gtrsim 1.8$ in the range $\tan\beta\lesssim 20$.  

\section{Supersymmetry breaking scale}
\label{susybreaking}
In the previous section we have computed, using the measured value of the Higgs mass, the value of the scale $\mathcal Q_0$ at which the MSSM matches with the Standard Model and the value of the parameter $m_2^2(\mathcal Q_0)$ which guarantees a correct electroweak Standard Model breaking at the scale $\mathcal Q_{EW}=m_H$. We are here making the conservative assumption (alas, consistent with present experimental data!) that only the SM states survive below the matching scale $\mathcal Q_0$. For large values of $\mathcal Q_0$ this amounts to assume a high-scale MSSM beyond $\mathcal Q_0$, in contradistinction with other possibilities, as those dubbed as split (or mini-split) supersymmetry. Using these tools we will now get information on the scale at which supersymmetry breaking is transmitted $\mathcal M$. 

As we have seen both $\mathcal Q_0$ and $m_2^2(\mathcal Q_0)$ are (for fixed values of the Standard Model parameters) functions of the MSSM parameters $\tan\beta$ and $X_t$ defined at the scale $\mathcal Q_0$: $\mathcal Q_0\equiv f_0(\tan\beta,X_t)$ and $m_2^2(\mathcal Q_0)\equiv f_2(\tan\beta,X_t)$. Now from the EWSB condition (\ref{minimum}) one can also compute $m_1^2(\mathcal Q_0)\equiv f_1(\tan\beta,X_t)$ as
\be
m_1^2(\mathcal Q_0)=m_2^2(\mathcal Q_0)\tan^2\beta+m^2(\mathcal Q_0)(\tan^2\beta-1)
\ee
so that both squared mass parameters $m_1^2$ and $m_2^2$ are fixed at the scale $\mathcal Q_0$ for fixed values of $\tan\beta$ and $X_t$. We will now define the scale at which supersymmetry is transmitted $\mathcal M$ as the scale at which
\be
m_1^2(\mathcal M)=m_2^2(\mathcal M)\ .
\label{unification}
\ee
where we are running the MSSM parameters from the scale $\mathcal Q=\mathcal Q_0$ to the scale $\mathcal Q=\mathcal M$ by using the two-loop RGE~\cite{Martin:1993zk}.
 Notice that this condition is rather generic in most models of supersymmetry breaking, as models based on gravity mediation or minimal gauge mediation, as well as in string constructions~\cite{Larry, Ibanez:2013gf ,Hebecker:2013lha}. 

As we are assuming that the effective theory below $\mathcal Q_0$ is just the Standard Model we are implicitly assuming that, at the matching scale the heavy Higgs $\mathcal H$ decouples, so that $m_{\mathcal H}(\mathcal Q_0)=\mathcal Q_0$. On the other hand the scale at which supersymmetry breaking is transmitted, given by (\ref{unification}), does have little dependence on the boundary conditions imposed for the rest of the supersymmetric spectrum. Thus we will next consider two generic situations.

\subsection{Bottom-up approach}
The most precise (and ideal) way by which the Standard Model will emerge as the low energy effective theory below the matching scale $\mathcal Q_0$ is when all supersymmetric particles are (approximately) degenerate at the decoupling scale~\footnote{Of course in practice there should be some spreading of supersymmetric masses over the scale $\mathcal Q_0$. A (more realistic) situation which will be studied in the next section.}. So we will here assume for all sfermions $(\widetilde f)$, Higgsinos (with mass $\mu$) and gauginos a degenerate mass at the matching scale $\mathcal Q_0$
\be
m_{\widetilde f}(\mathcal Q_0)=M_i(\mathcal Q_0)=\mu(\mathcal Q_0)=\mathcal Q_0\quad (i=1,2,3)
\label{decbottomup}
\ee
We will leave $X_t(\mathcal Q_0)$ [and consequently the mixing $A_t(\mathcal Q_0)$] and $\tan\beta(\mathcal Q_0)$ as free parameters in the plots. 

\begin{figure}[htb]\hspace{5mm}
\begin{center}
\includegraphics[width=81.5mm,height=70mm]{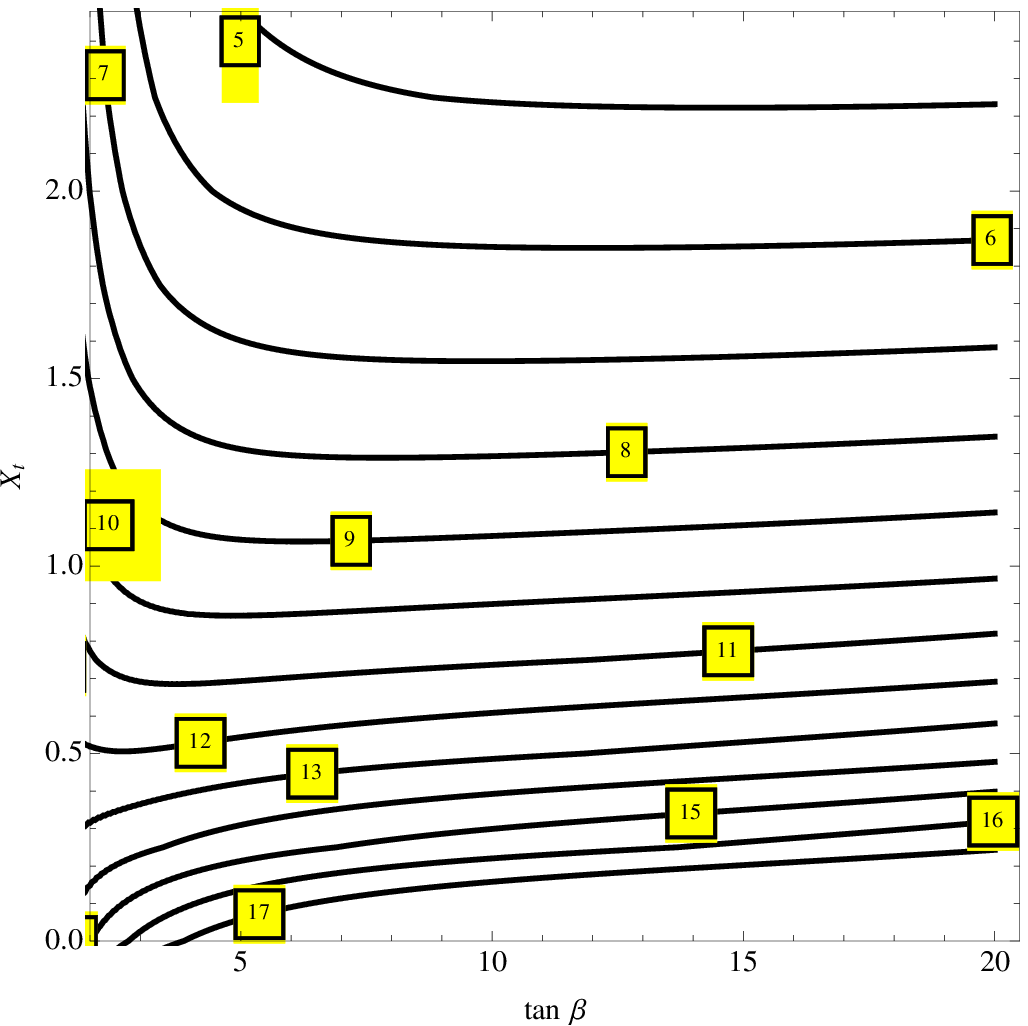}
\includegraphics[width=81.5mm,height=70mm]{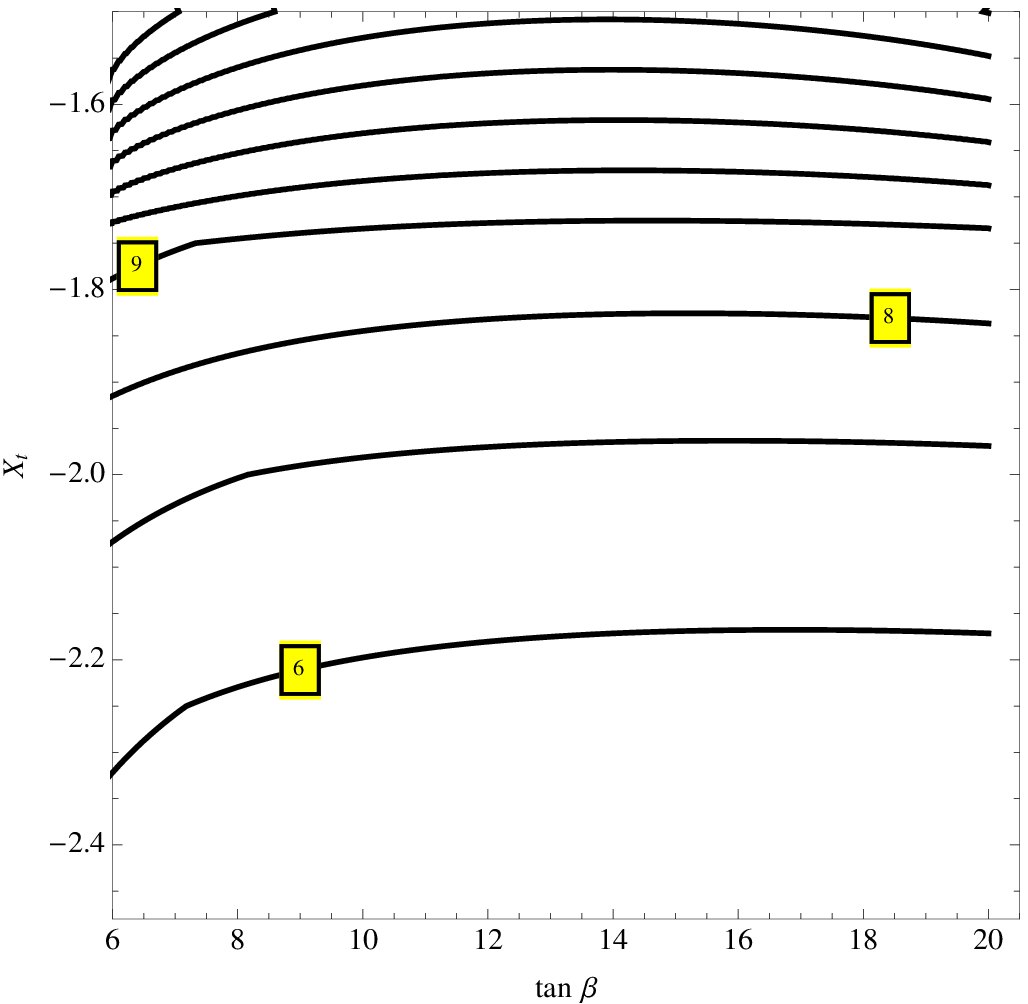}
\end{center}
\caption{\it  Contour lines of constant $\log_{10}[\mathcal M/{\rm GeV}]$ in the $(\tan\beta,X_t)$ plane for $X_t\geq 0$ (left panel) and  $X_t<0$ (right panel).}
\label{Mcontours}
\end{figure}

Note that by imposing the matching scheme in Eq.~(\ref{decbottomup}) the merging between the SM and the MSSM happens at the scale $\mathcal Q_0$ and the running from the low scale $\mathcal Q_0$ to the high scale $\mathcal M$ can be done straightforwardly using the two-loop MSSM RGE and  the boundary conditions (\ref{decbottomup}). This is shown in the left panel (right panel) of Fig.~\ref{Mcontours}
where we plot contour lines of constant $\log_{10}(\mathcal M/{\rm GeV})$ in the $(\tan\beta,X_t)$ plane for the central value of the top quark mass and positive (negative) values of the parameter $X_t$.

We can see from the left panel of Fig.~\ref{Mcontours} that having supersymmetry breaking transmission at high scale requires both large values of $\tan\beta$ and small and positive values of the mixing $X_t$. For example for values of $\mathcal M$ of the order of the unification scale $\mathcal M\simeq 10^{16}$ GeV one requires $\tan\beta\gtrsim 3$ and $X_t\lesssim 0.3$.  Moreover for large values of $\tan\beta$ the value of $\mathcal M$ depends almost uniquely on the mixing $X_t$. For example even for $\tan\beta\simeq 20$ the scale at which supersymmetry is broken can go down to values as low as $\mathcal M\sim10^5-10^6$ GeV for values of the mixing $X_t\simeq 2$. On the other hand for low values of $\tan\beta$ and large values of $X_t$ there is small dependence on the mixing. 
As we can see from the left panel of Fig.~\ref{Mcontours} for values $X_t\simeq 0$ we can get values of $\mathcal M$ as large as $M_P$. For negative values of $X_t$ the value of $\mathcal M$ grows quickly to trans Planckian values and rapidly disappears as there is no solution to the Eq.~\ref{unification}. A solution appears again for values $X_t\simeq -1.5$ for which we have again values of $\mathcal M\simeq M_P$, and again the values of $\mathcal M$ decrease when we increase the absolute value of $X_t$ as we have shown in the right panel of Fig.~\ref{Mcontours}. 

\begin{figure}[htb]
\begin{center}
\includegraphics[width=81.5mm]{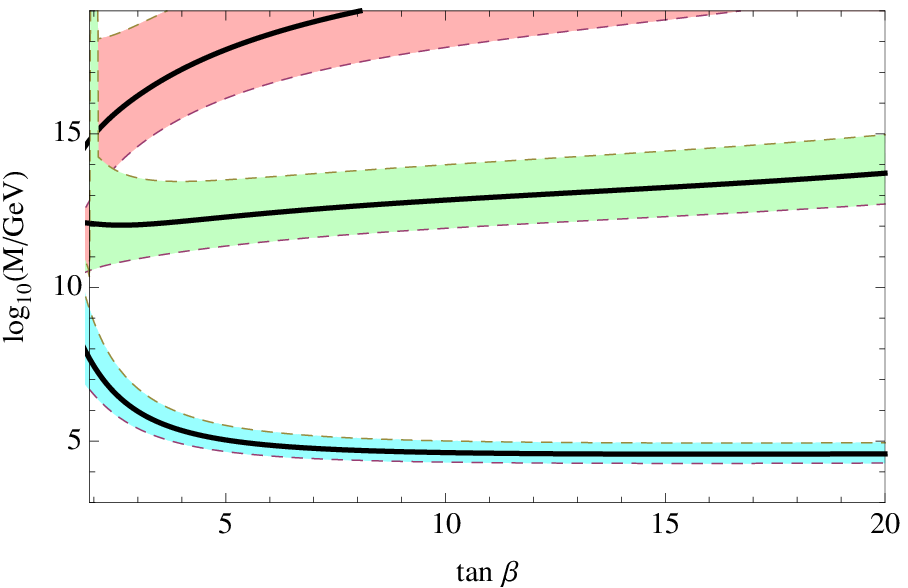}
\includegraphics[width=81.5mm]{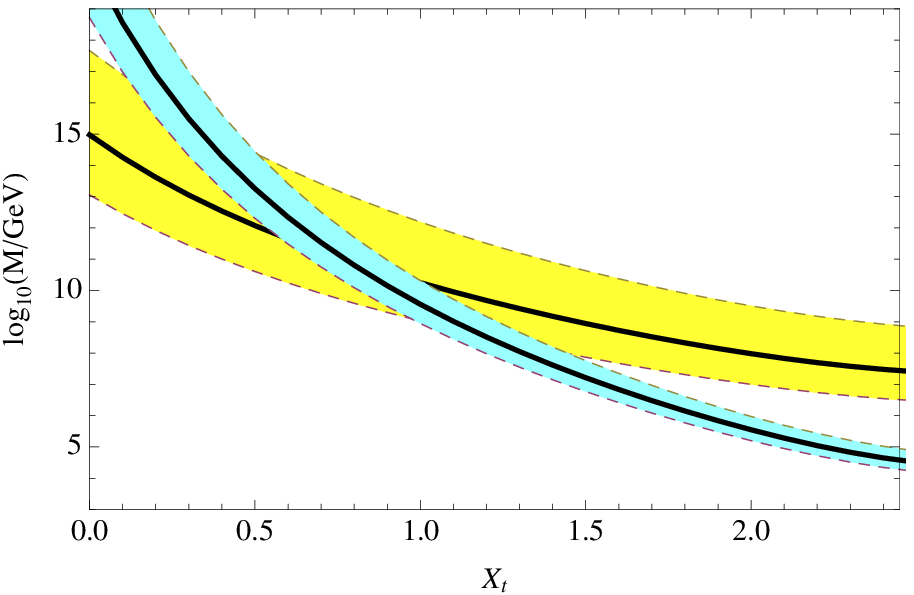}
\end{center}
\caption{\it Left panel: Plot of $\log_{10}[\mathcal M/GeV]$ as a function of $\tan\beta$ for $X_t=$0 (upper band), $X_t=$0.5 (central band) and $X_t=\sqrt{6}$ (lower band). The width of bands corresponds to the experimental error $\Delta \overline{m}_t=\pm 2$ GeV. Right panel:  Plot of $\mathcal M$ as a function of $X_t$ for $\tan\beta=2$ (upper band) and $15$ (lower band). }
\label{Mplot}
\end{figure}

Of course, as it was the case of the matching scale $\mathcal Q_0$, the scale at which supersymmetry is transmitted $\mathcal M$ is affected by the experimental error in the determination of the top quark mass $\Delta\overline{m}_t$. This effect is shown numerically in Fig.~\ref{Mplot}. 
We plot in the left panel of Fig.~\ref{Mplot} $\log_{10}(\mathcal M/GeV)$ as a function of $\tan\beta$ for different values of the mixing $X_t=$0, 0.5, and $\sqrt{6}$ for the values of the $\overline{\rm MS}$ top quark mass $\overline{m}_t(m_t)=163.5\pm 2$ GeV. This effect is mainly inherited from the uncertainty in the determination of the matching scale $\mathcal Q_0$, which explains why the effect is larger for $\tan\beta=1$.  Similarly the plot of $\log_{10}(\mathcal M/GeV)$ as a function of $X_t$ for fixed values of $\tan\beta=$2 and 15, is shown in the right panel of Fig.~\ref{Mplot} where we can also see that the uncertainty in the determination of $\mathcal M$ decreases with increasing values of $\tan\beta$.

\subsection{Top-down approach}
In the previous section we have assumed that all supersymmetric particles exactly decouple at the matching scale $\mathcal Q_0$, by which we were assuming a degenerate spectrum at this scale. Of course this is not the generic case in (realistic) models of supersymmetry breaking which provide some pattern of masses at the scale $\mathcal M$. These masses run, with the MSSM RGE, from the scale $\mathcal M$ to $\mathcal Q_0$ and thus they decouple at the scale $\sim\mathcal Q_0$ with different thresholds. 

In this section we will consider different supersymmetric spectra, for which the scale at which supersymmetry breaking is transmitted and the matching scale with the Standard Model satisfy the general values which have been obtained in the previous section: in particular they are consistent with electroweak symmetry breaking with a Higgs mass of 126 GeV. We will not commit ourselves to any particular mechanism of supersymmetry breaking but instead will consider generic pattern of supersymmetric spectra at the scale where supersymmetry breaking is transmitted, which can arise from different mechanisms. In particular we will consider two classes of models, which are simply particular examples while many others can be easily found and studied: 
\begin{itemize}
\item
Models with universal soft parameters, typical of gravity mediated-like models, although not necessarily arising from gravity mediation.
\item
Gauge mediated models, where the values of supersymmetric parameters satisfy, at the scale $\mathcal M$, typical ratios provided by gauge mediation.
\end{itemize}

\subsubsection{Universal soft parameters}

In this section we are going to consider some universal soft breaking parameters at the scale at which supersymmetry breaking is transmitted $\mathcal M$. In particular we will assume the rather general pattern
\be
m_{\widetilde Q_3}(\mathcal M)=m_{\widetilde U_3^c}(\mathcal M)=m_{\widetilde D_3^c}(\mathcal M)\equiv m_0,\quad M_i(\mathcal M)\equiv m_{1/2},\quad m_1(\mathcal M)=m_2(\mathcal M)
\label{universal}
\ee
by which all third generation squarks~\footnote{Third generation sleptons as well as first and second generation sfermions do not play any role in the RGE and thus their values decouple from the present problem.} are degenerate at the scale $\mathcal M$, as well as the three gauginos and the two MSSM Higgs doublets. We have then considered the common masses $m_0$ and $m_{1/2}$ as free parameters only subject to the constraint of getting a correct electroweak symmetry breaking.

\begin{figure}[htb]
\begin{center}
\includegraphics[width=89.5mm]{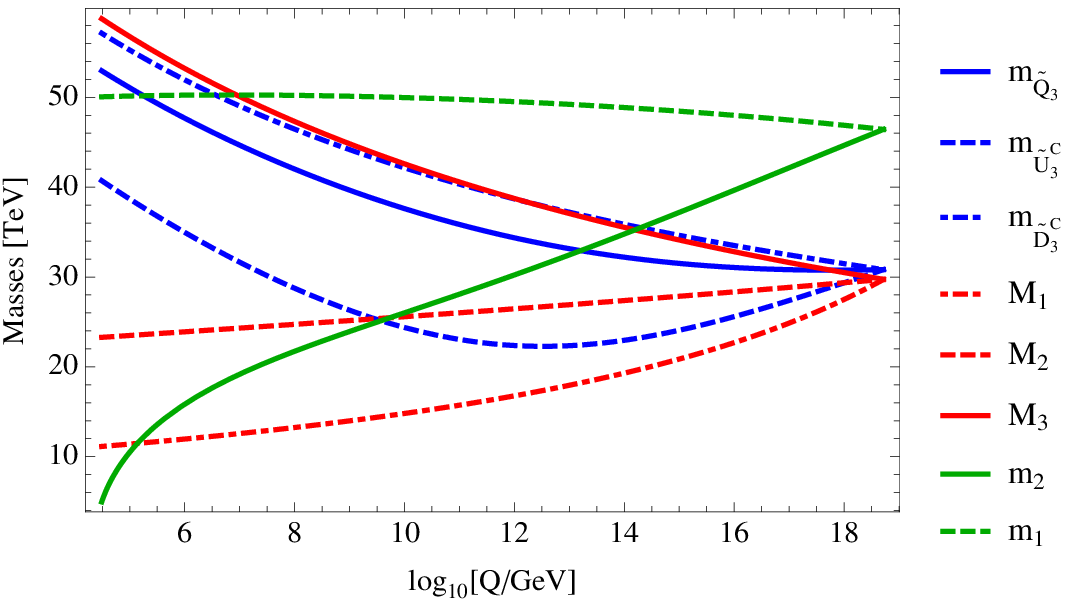}
\hspace{-1mm}\includegraphics[width=74.5mm]{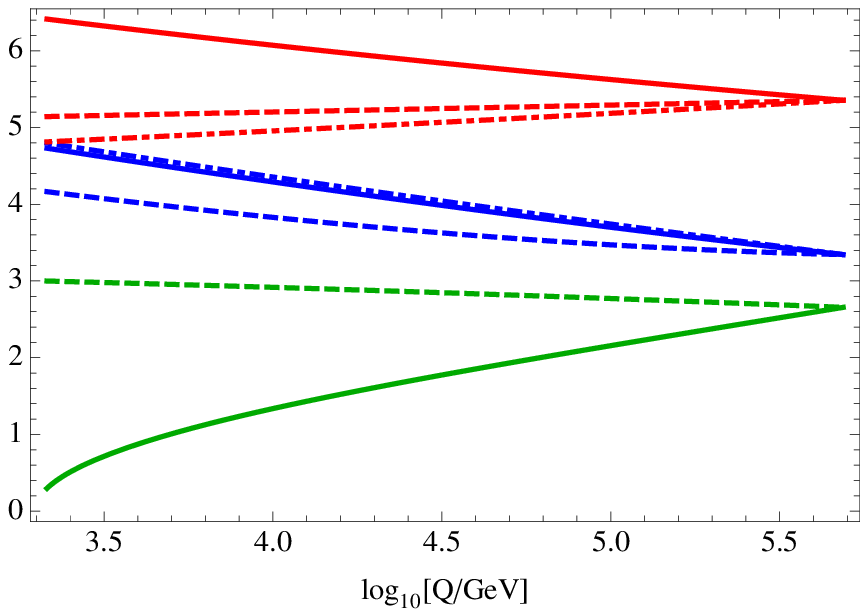}
\end{center}
\caption{\it RGE running between $\mathcal M$ and $\mathcal Q_0$ of the supersymmetric spectrum for the case $\tan\beta=10$, $X_t=0$ (left panel) and $X_t=2$ (right panel) with universal boundary conditions.}
\label{universalfig}
\end{figure}

We have considered in Fig.~\ref{universalfig} two generic models which correspond to $\tan\beta=10$,
and $X_t=0$ (left panel) and $X_t=2$ (right panel). As for the case of $X_t=0$ a quick glance at the left panel of Fig.~\ref{Q0contours} shows that the matching scale is $\mathcal Q_0\sim 100$ TeV while from Fig.~\ref{Mcontours} the scale where supersymmetry breaks is $\mathcal M\sim 2\times 10^{18}$ GeV. Also from the right panel of Fig.~\ref{Q0contours} we see that the breaking is not radiative in the sense that $m_2^2(\mathcal Q_0)>0$ and indeed from Fig.~\ref{m2} we can see that, according with the correct electroweak symmetry breaking, $m_2(\mathcal Q_0)\simeq 3$ TeV. As we can see from the left panel of Fig.~\ref{universalfig} the values for the common squark and gaugino masses which fit these conditions are: $m_0\simeq m_{1/2}\simeq 30$ TeV. Also the value of $X_t=0$ at the matching scale $\mathcal Q_0$ translates into the mixing $A_t(\mathcal M)\simeq 1.7 \,m_0$. Notice that, as the value of $\mathcal M$ is around the Planck scale, this scenario could arise in models where supersymmetry breaking is transmitted by gravitational interactions.

If we now increase the value of $X_t$, as in the right panel of Fig.~\ref{universalfig}, in which $X_t=2$, then looking again at Fig.~\ref{Q0contours} we see that the matching scale is $\mathcal Q_0\sim 1$ TeV and the electroweak breaking is (almost) radiative as $m_2(\mathcal Q_0)\sim 100$ GeV. Likewise, from Fig.~\ref{Mcontours}, the scale at which supersymmetry is broken is $\mathcal M\sim 5\times 10^5$ GeV. Here we can see a general phenomenon by which the scale where supersymmetry breaking is transmitted (i.e. the scale of unification of $m_1$ and $m_2$) strongly goes down when the mixing increases if we fix the correct conditions for electroweak breaking. The reason is the contribution of the mixing to the RGE as
\be
\beta_{m_2^2}=\frac{3h_t^2}{4\pi^2}A_t^2+\cdots.
\ee 
To prevent electroweak breaking at high scale ($\mathcal Q\gg\mathcal Q_0$) we then let the scale $\mathcal M$ go down. For the same reason we need gauginos heavier than squarks as the former ones contribute with negative sign to $\beta_{m_2^2}$. As we can see in the right panel of Fig.~\ref{universalfig} this condition translates into $m_0\simeq 3.3$ TeV and $m_{1/2}\simeq 5.3$ TeV while at the matching scale $\mathcal Q_0$ all the supersymmetric spectrum is in the interval $3-6$ TeV.

\subsubsection{Gauge mediated models}
In this section we will apply the previous results to the particular case in which supersymmetry breaking is transmitted to the observable sector by gauge interactions (GMSB). We will assume in particular the minimal GMSB model whose main features we now summarize.

Supersymmetry is broken, in a hidden sector, by a spurion chiral superfield $X=F\theta^2$ which is coupled to a set of pairs, $\Phi_i+\overline\Phi_i$, of messenger fields, in vector like ${\bf r}+\overline{\bf r}$ representations of the gauge group  with the superpotential $W=\sum_{i}\Phi_i \{\lambda_i X+M_i\}\overline \Phi_i$.

Gauginos acquire a Majorana mass, by one loop diagrams, given by~\cite{Giudice:1998bp}
\be
M_a(\mathcal M)=\frac{\alpha_a(\mathcal M)}{4\pi} \Lambda_G, \qquad
\Lambda_G\simeq\sum_i n_i\frac{\lambda_i F}{M_i}=N\frac{F}{M}
\label{gauginomass}
\ee
where $n_i$ is the Dynkin index for the pair $\Phi_i+\overline\Phi_i$~\footnote{We are using a normalization where $n_{SU({\bf N})}=1$ for the ${\bf N}+\overline{\bf N}$ representation of $SU({\bf N})$, $n_{U(1)}=6Y^2/5$, and $\alpha_1$ is the $U(1)$ gauge coupling which satisfies the unification condition $\alpha_a(M_{GUT})=\alpha_{GUT}$.},  and $N=\sum_i n_i$. For the last equality of Eq.~(\ref{gauginomass}) we are assuming universal messenger masses as $M_i\equiv \lambda_iM$ (for $\forall i$). Likewise supersymmetric scalars (squarks and sleptons) acquire soft breaking squared masses through two loop diagrams as
\be 
m^2_{\widetilde f\,}(\mathcal M)=2\sum_{a}C_a^{\widetilde f}\,\frac{\alpha_a^2(\mathcal M)}{16\pi^2}\Lambda_S^2,\quad \Lambda_S^2=\sum_i n_i\frac{(\lambda_i F)^2}{M_i^2}=N\frac{F^2}{M^2}
\label{scalarmass}
\ee
where $C_a^{\widetilde f}$ is the quadratic Casimir of the representation to which $\widetilde f$ belongs in the group $G_a$~\footnote{We are using a normalization where for $SU(3)$ triplets, $C_3=4/3$, for $SU(2)_L$ doublets, $C_2=3/4$, and $C_1=3 Y^2/5$. In all cases $C_a=0$ for gauge singlets.}, and again for the last equality of Eq.~(\ref{scalarmass}) we are assuming universal messenger masses. In fact for the case of universal messenger masses the ratio $\Lambda_G^2/\Lambda_S^2=N$ is given by the number of messengers, however in more general cases (which can arise e.g.~for several $X$ fields overlapping with the Goldstino field) one can treat $\Lambda_G$ and $\Lambda_S$ as free parameters. The soft breaking parameter $A_t$ is not generated at one loop so we will fix it as $A_t(\mathcal M)=0$ and will let it to develop at the scale $\mathcal Q_0$ by the MSSM RGE running, which is equivalent to a two loop effect.

 \begin{figure}[htb]
\begin{center}
\includegraphics[width=89.5mm]{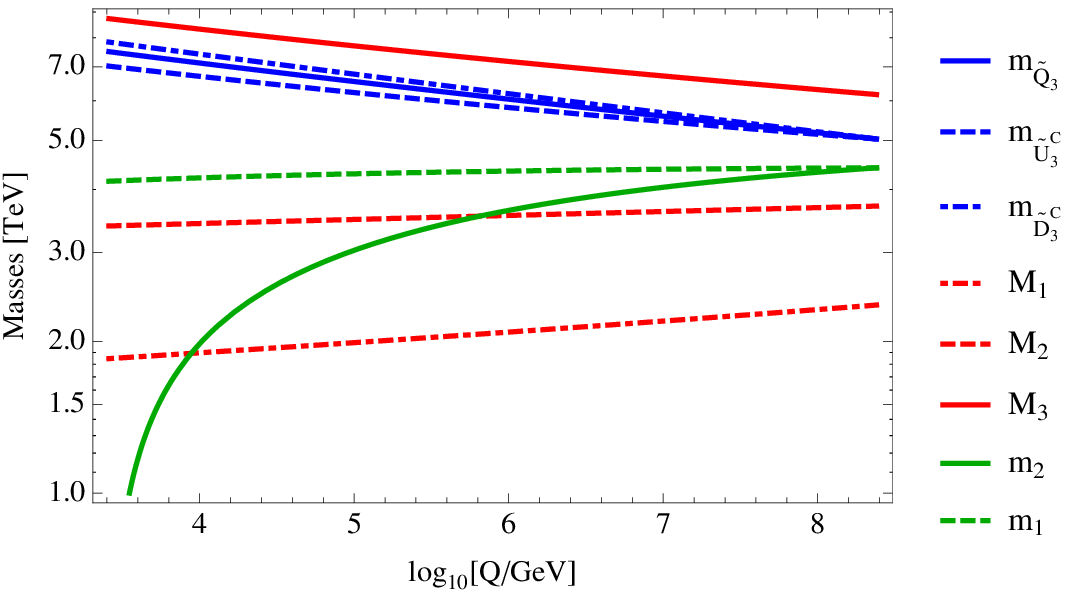}
\hspace{-5mm}\includegraphics[width=75.5mm]{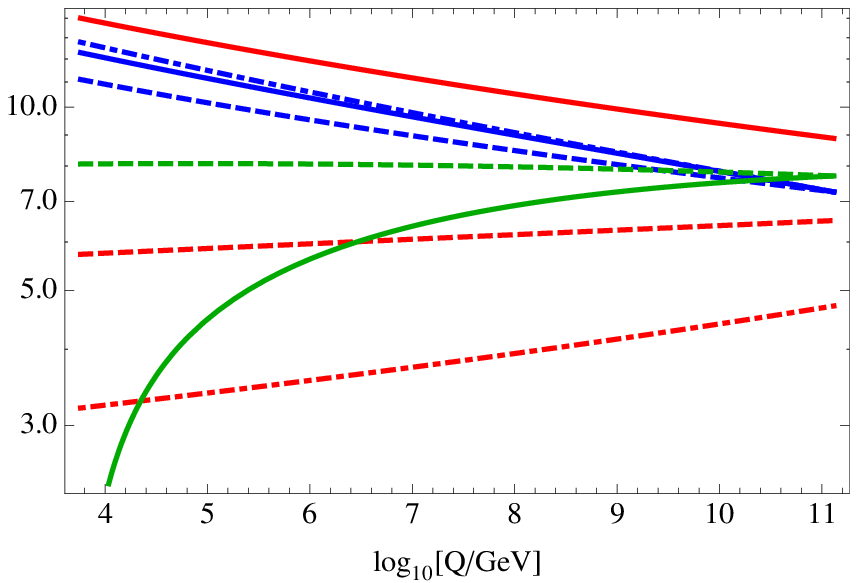}
\end{center}
\caption{\it RGE running between $\mathcal M$ and $\mathcal Q_0$ of the supersymmetric spectrum for the case $A_t(\mathcal M)=0$ and $\tan\beta=15$ (left panel) and $\tan\beta=8$ (right panel) with gauge mediated boundary conditions.}
\label{gaugemediation}
\end{figure}

In Fig.~\ref{gaugemediation} we are presenting two typical cases where GMSB is consistent with the conditions imposed by electroweak breaking for a 126 GeV Higgs mass. The case $\tan\beta=15$ is presented in the left panel and $\tan\beta=8$ in the right panel. In both cases we have fixed $\Lambda_G=2\Lambda_S$ which corresponds to four messengers, $N=4$, in minimal GMSB models. Both cases are, as we will see, consistent with perturbative unification. 

For the case $\tan\beta=15$ in the left panel of Fig.~\ref{gaugemediation} we get $\Lambda_G\simeq 1.4\times 10^6$ GeV, $\mathcal M\simeq 3\times 10^{8}$ GeV, and the scale of supersymmetry breaking $\sqrt{F}\simeq 10^7$ GeV while the expansion parameter $F/\mathcal M^2\simeq 10^{-3}$ is small, and the gravitino mass is $m_{3/2}\simeq 20$ keV. Notice that $m_{H_i}^2(\mathcal M)<m_{\widetilde Q}^2(\mathcal M)$ although $m_i^2(\mathcal M)>m_{\widetilde Q}^2(\mathcal M)$ because of the contribution of $\mu^2$ in $m_i^2$. This case is perfectly consistent with perturbative unification and the messengers change the value of the gauge couplings at the unification scale by $\delta\alpha_{GUT}^{-1}\simeq -11$. Even if $A_t(\mathcal M)=0$ a nonzero (and negative) value is generated at the scale $\mathcal Q_0$ such that $X_t\simeq -1.8$.

For the case shown in the right panel of Fig.~\ref{gaugemediation} that corresponds to $\tan\beta=8$ we get the following values of the parameters: $\Lambda_G\simeq 2 \times 10^6$ GeV, $\mathcal M\simeq  10^{11}$ GeV, $\sqrt{F}\simeq 3 \times 10^8$ GeV with the expansion parameter  $F/\mathcal M^2\simeq 4\times 10^{-6}$ and $m_{3/2}\simeq 20$ MeV. This case is also consistent with perturbative unification with a value of the gauge couplings at the unification scale and the messengers change the value of the gauge couplings at the unification scale by $\delta\alpha_{GUT}^{-1}\simeq -8$. Similarly a nonzero negative value of $X_t$ is generated as $X_t\simeq-1.6 $.

\section{Conclusions}
\label{conclusion}
The Standard Model is consistent with all present experimental data including the recent measurements of the Higgs mass and its couplings to gauge bosons and fermions. By the same token experimental data are putting bounds on possible BSM physics whose aim is to solve the SM hierarchy problem, i.e.~to understand the hierarchy $\mathcal Q_{EW}/M_P\simeq 10^{-16}$, or equivalently the stability of the electroweak vacuum. Even if no hint of new physics have been found by the LHC (and even in the case that no positive signal be found by the future LHC runs) still the stability of the big hierarchy between the LHC energy scale and the Planck scale $(\mathcal Q_{LHC}/M_P\simeq 10^{-14})$ requires a theoretical explanation, although theories aiming to  explaining the little hierarchy ($\mathcal Q_{EW}/\mathcal Q_{LHC}\simeq 10^{-2}$) do not receive support from the experimental side. On the other hand the paradigm of theories solving the hierarchy problem is supersymmetry, which has roots in superstring theories for which low-scale supersymmetry is not an essential ingredient.

So a possible attitude (that we have adopted in this paper) is to assume that supersymmetry is solving the big hierarchy problem from $\mathcal Q_{LHC}/M_P$ (which amounts to a fine-tuning of one part in $10^{28}$!) but perhaps not necessarily the little hierarchy problem from $\mathcal Q_{EW}/\mathcal Q_{LHC}$ (which amounts to a fine-tuning of around one part in ten thousand) and see what the present data are telling us about the parameters of the supersymmetric theory.

Using then the measured value of the Higgs mass and imposing the conditions for electroweak symmetry breaking we can obtain information on the scale of supersymmetric parameters $(\mathcal Q_0)$ and the conditions on how the supersymmetric theory triggers electroweak breaking. Moreover by making the mild assumption that the mass parameters of both Higgs bosons unify at the scale at which supersymmetry breaking is transmitted $\mathcal M$ we can obtain rather general information on the latter, as we have described throughout this paper. In models where the former assumption on the Higgs bosons mass at $\mathcal M$ is not fulfilled the conditions should be accordingly modified.

Our analysis just reflects the present experimental situation concerning the Higgs discovery and the non-observation of any supersymmetric particle in the LHC7 and LHC8 runs. In the future, when the LHC13-14 run will start in 2015, it might happen that supersymmetric signals are found or that they are not. In both cases the present analysis should be correspondingly constrained. In case where supersymmetric signals are found, they would give information about our energy scale $\mathcal Q_0$ which in turn will give indirect information about the scale at which supersymmetry breaking is transmitted $\mathcal M$. In the other case, in which supersymmetric signals not be found at the LHC13-14, the data will put a lower bound on the scale $\mathcal Q_0$ by which also the scale $\mathcal M$ will be correspondingly constrained, suggesting that perhaps we will need a higher energy collider to uncover BSM physics as the HE-LHC (at 33 TeV) \& VHE-LHC (at 100 TeV)~\cite{Koratzinos:2013chw}.

\section*{Acknowledgments}
AD and MQ would like to thank partial support from the National Science Foundation under Grant No. PHY11-25915. The work of AD was partially supported by the National Science Foundation under Grant No. PHY12-15979. The work of MQ was supported in part by the European Commission under the ERC Advanced Grant BSMOXFORD 228169, by the Spanish Consolider-Ingenio 2010 Programme CPAN (CSD2007-00042), and by CICYT-FEDER-FPA2011-25948.

 \end{document}